\documentclass[seceq]{ptptex}
\usepackage{epsfig}

\title{
Cluster Monte Carlo Algorithms for Dissipative Quantum Systems%
}

\author{
Philipp \textsc{Werner} and Matthias \textsc{Troyer}%
}

\inst{
Institut f\"ur theoretische Physik, ETH Z\"urich, CH-8093 Z\"urich, Switzerland
}

\abst{
We review efficient Monte Carlo methods for simulating quantum systems which couple to a dissipative environment. A brief introduction of the Caldeira-Leggett model and the Monte Carlo method will be followed by a detailed discussion of cluster algorithms and the treatment of long-range interactions. Dissipative quantum spins and resistively shunted Josephson junctions will be considered.
}

\begin{document}

\maketitle

\section{Introduction}

\noindent

Dissipation and decoherence in
quantum systems have
important practical implications, from stabilizing
superconductivity in granular materials, to the loss of
information stored in qubits.
 In all the models we study,  energy dissipation is introduced by coupling a quantum mechanical degree of freedom to a bath of suitably chosen harmonic oscillators. The degrees of freedom of this bath can be integrated out analytically, which leads to an effective action with long-range interactions in imaginary time. 
In this article we will review and present a number of recently developed efficient cluster Monte Carlo algorithms for various dissipative quantum systems.
These algorithms have been used to study spin chains with dissipation coupling to the site variables and resistively shunted Josephson junctions \cite{ising, smt, xy, ebox, junction, twojunction, diss}. We will also present relevant theoretical background for these models and mention the open questions which have been addressed in the numerical investigations.    

\section{Dissipative quantum models}
\subsection{Caldeira-Leggett model}
While the concept of coupling a quantum mechanical degree of freedom to an environment goes back to the early days of quantum mechanics, it was the work of Caldeira and Leggett \cite{C&L_letter, Caldeira} which pointed out the important consequences of dissipation at a time when experimental results on macroscopic quantum tunneling became available. 
To introduce their model we will consider a classical system whose dissipative dynamics in terms of some coordinate $q$ is described in the absence of external forces by the phenomenological equation of motion
\begin{equation}
M\ddot q(t) + \eta \dot q(t) + [V'(q)](t) = 0, \label{damped_motion}
\end{equation}
where $\eta$ denotes the friction coefficient. Because the frictional force is proportional to $\dot q$, one calls the dissipation ``Ohmic''. 

The equation of motion (\ref{damped_motion}) can be obtained from a microscopic model by 
coupling the system linearly in the coordinates
to an appropriately chosen
set of harmonic oscillators, that is, 
from the Lagrangian
\begin{equation}
L = \frac{M}{2}\dot q^2 - V(q) +
\frac{1}{2}\sum_\alpha\left\{m_\alpha\dot x_{\alpha}^2 -
m_\alpha\omega_\alpha^2 \left(x_\alpha -
\frac{c_\alpha}{m_\alpha\omega_\alpha^2}q\right)^2 \right\},
\label{1.3}
\end{equation}
where $x_\alpha$, $m_\alpha$, $\omega_\alpha$ and $c_\alpha$ denote the oscillator
positions, masses, frequencies and coupling strengths, respectively. 
Defining the spectral density
\begin{equation}
J(\omega) :=
\frac{\pi}{2}\sum_\alpha\frac{c_\alpha^2}{m_\alpha\omega_\alpha}
\delta(\omega - \omega_\alpha ), \label{1.9}
\end{equation}
we can reproduce
the phenomenological equation (\ref{damped_motion}) by setting
\begin{equation}
J(\omega) = \eta\omega,\label{1.14}
\end{equation}
for $\omega\ge 0$ \cite{Caldeira, Weiss, diss}. In any real physical system, the harmonic oscillators representing the environment
will have finite frequencies, so the linear relationship
(\ref{1.14}) must be cut off at some value $\omega_0$.

The partition function of the quantum mechanical system can be
obtained from the classical Euclidean Lagrangian 
\begin{equation}
L^E = \frac{M}{2}\dot q^2 + V(q) +
\frac{1}{2}\sum_\alpha\left\{m_\alpha\dot x_{\alpha}^2 +
m_\alpha\omega_\alpha^2 \left(x_\alpha -
\frac{c_\alpha}{m_\alpha\omega_\alpha^2}q\right)^2 \right\}, \label{2.1}
\end{equation}
by the path integral formalism \cite{Caldeira, Weiss, diss}. Integrating out the  harmonic oscillator degrees of freedom leads to an effective action
\begin{eqnarray}
S_{\text{eff}}^E[q] &=&\int_0^\beta d\tau \Big[\frac{M}{2}\dot
q^2(\tau)+V(q(\tau))\Big] +
\frac{\eta}{4\pi}\int_0^\beta\!\!\int_0^\beta d\tau d\tau'
\Big(\frac{\pi}{\beta}\Big)^2\frac{(q(\tau)-q(\tau'))^2}{(\sin(\frac{\pi}{\beta}|\tau-\tau'|))^2}.
\hspace{0mm}\nonumber\\ \label{2.27}
\end{eqnarray}

In analytical calculations, but also for certain algorithms discussed in chapter 3, it is useful to express the damping term (proportional to $\eta$) in Fourier space. In the limit $\beta\rightarrow\infty$ and for asymmetric Fourier transforms one obtains
\begin{equation}
S_{\text{damp}} =
\frac{\eta}{4\pi}\int_{-\infty}^\infty d\omega|\omega||q(\omega)|^2,\label{3.4}
\end{equation}
which is local in the frequency $\omega$, in contrast to the imaginary-time integral (\ref{2.27}).

\subsection{Dissipative Quantum Spin Chains}

The effective action for a single $O(n)$ spin with on-site dissipation is obtained from
Eq.~(\ref{2.27}) by choosing an $n$-component vector $\vec q$ as the coordinate which couples to the environment and a double-well potential
\begin{equation}
V(q) = \frac{s}{2}q^2 + \frac{u}{24}q^4 \label{V_double_well}
\end{equation}
with minima at $|\vec q|=1$. Coupling the spins spatially through nearest neighbor bonds of the form $-K\vec q_j\cdot \vec q_{j+1}$ then yields the action for the spin chain
\begin{eqnarray}
S &=& \sum_{j=1}^{N_x}\int_0^\beta d\tau\left[\frac{M}{2}(\partial_\tau
\vec q_j)^2 -K\vec q_j\cdot \vec q_{j+1} + \frac{s}{2}q_j^2 +
\frac{u}{24}q_j^4 \right]\nonumber\\
&& +
\frac{\eta}{4\pi}\sum_{j=1}^{N_x}\int_0^\beta\!\!\int_0^\beta d\tau d\tau'
\Big(\frac{\pi}{\beta}\Big)^2\frac{(\vec q_j(\tau)-\vec q_j(\tau'))^2}{(\sin(\frac{\pi}{\beta}|\tau-\tau'|))^2}.\label{3.15}
\end{eqnarray}
The special cases $n=1$ (Ising spin chain) and $n=2$ (XY-spin chain) will be discussed in chapter 4.

To simplify the calculations, we will discretize the action (\ref{3.15}) in imaginary time, noting that the cutoff of the linear spectral density $J(\omega)$ at $\omega_c$ would anyhow lead to modifications in $S_\text{\text{damp}}$ at small $\tau-\tau'$. We use a Trotter number $N_\tau$, corresponding to a discretization step $\Delta \tau=\beta/N_\tau$. Furthermore we employ a ``hard spin'' approximation and choose the potential (\ref{V_double_well}) such that the spin states at any time can be described as $O(n)$ vectors $\vec\sigma$ of norm $|\vec\sigma|=1$. Introducing the the coefficients $\Gamma=M/\Delta\tau$ and $\alpha=\eta/\pi$, the discretize action becomes
\begin{equation}
S = -\sum_{j=1}^{N_x}\sum_{k=1}^{N_\tau}\left[K\vec \sigma_{j,k}\cdot \vec \sigma_{j+1,k}+\Gamma \vec \sigma_{j,k}\cdot \vec \sigma_{j,k+1}\right]-\alpha\sum_{j=1}^{N_x}\sum_{k<k'}\Big(\frac{\pi}{N_\tau}\Big)^2\frac{\vec \sigma_{j,k}\cdot \vec \sigma_{j,k'}}{(\sin(\frac{\pi}{N_\tau}|k-k'|))^2}.\label{spin_lattice},
\end{equation}
which can be simulated as a two-dimensional system of classical $O(n)$-spins with short-range interactions in the ``space"- and long-range interactions in the ``imaginary time" direction.

\subsection{Resistively shunted Josephson junction}

\begin{figure}[t]
\centering
\includegraphics [angle=0, width= 0.4\textwidth] {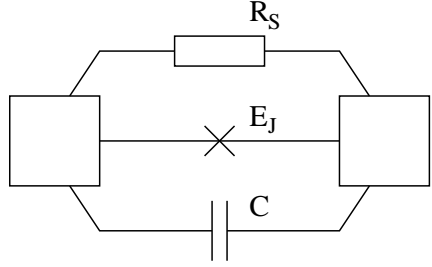}
\caption{Illustration of a resistively shunted Josephson junction. It consists of two superconducting islands connected by a Josephson junction of coupling energy $E_J$ and capacitance $C$ and a parallel resistor $R_s$ through which electrons can flow}.
\label{junction}
\end{figure}

Another example of a dissipation coupled quantum system is
a Josephson junction with Josephson coupling energy $E_J$ and capacitance $C$, which is shunted by an Ohmic resistor $R_s$, as illustrated in Fig.~\ref{junction}. While Cooper pairs can tunnel through the junction, which consists of a thin layer of insulating material or a constriction in the superconductor, electrons can flow through the resistor and dissipate energy. Let $\phi_l$ and $\phi_r$ denote the phases of the superconducting order parameter on the left and right island, respectively, and $\phi=\phi_l-\phi_r$ the phase difference across the junction. This phase difference is related to the voltage drop $V$ across the junction and the super-current $I_s$ by the Josephson relations ($h/2e$ is the flux quantum) \cite{Josephson}
\begin{eqnarray}
V &=& \frac{\hbar}{2e}\dot\phi,\label{V}\\
I_s &=& I_c\sin\phi,\label{I_s}
\end{eqnarray}
where $I_c$ denotes the critical current, which is related to the coupling energy by 
\begin{equation}
E_J=\frac{\hbar I_c}{2e}.\label{E_J}
\end{equation}
The dynamics of such a device can be determined from a balance of currents, which for zero external current reads
\begin{eqnarray}
0&=&I_\text{displacement}+I_\text{shunt}+I_\text{supra}= C\dot V + \frac{V}{R_s} + I_c\sin\phi\nonumber\\
&=&C\frac{\hbar\ddot\phi}{2e}+\frac{1}{R_s}\frac{\hbar\dot\phi}{2e}+ \frac{2e E_J}{\hbar}\sin\phi.\label{balance}
\end{eqnarray}
If we identify $\frac{\hbar\phi}{2e}$ with a coordinate $q$ and introduce the potential
\begin{equation}
U(q) = -E_J\cos(\phi(q))=-E_J\cos((2e/\hbar)q),\label{potential}
\end{equation}
then (\ref{balance}) becomes
\begin{equation}
C\ddot q + \frac{1}{R_s}\dot q + U'(q) =0,\label{simple_balance}
\end{equation}
which is just equation (\ref{damped_motion}) with the substitutions $M\rightarrow C$ and $\eta\rightarrow \frac{1}{R_s}$. According to the recipe of  Caldeira and Leggett outlined in section 2.1, the dissipation in the shunt resistor can be reproduced by coupling $q$ to an Ohmic heat bath, which is integrated out and leads to the effective action (\ref{2.27}). Expressing $q$, $M$ and $\eta$ again in terms of the original variables $\phi$, $C$ and $R_s$, one finds the action for the resistively shunted Josephson junction. Introducing furthermore the effective charging energy of the junction $E_C=\frac{e^2}{2C}$ (which sets the energy scale) and the quantum of resistance $R_Q = \frac{h}{4e^2}$, the action becomes (setting $\hbar=1$)
\begin{eqnarray}
S[\phi] &=& \int_0^\beta d\tau \left[\frac{1}{16E_C}\Big(\frac{d\phi}{d\tau}\Big)^2-E_J\cos(\phi)\right]\nonumber\\
&&+\frac{1}{8\pi^2}\frac{R_Q}{R_s}\int_0^\beta \int_0^\beta d\tau d\tau' \frac{(\pi/\beta)^2(\phi(\tau)-\phi(\tau'))^2}{\sin((\pi/\beta)(\tau-\tau'))^2}.\hspace{5mm}
\label{junction_action}
\end{eqnarray}  

Resistively shunted Josephson junctions at zero temperature undergo a super-conductor-to-metal transition if the shunt resistance $R_s$ is increased beyond the critical value $R_Q$. 
This dissipative phase transition was first predicted by Schmid~\cite{Schmid} and Bulgadaev~\cite{Bulgadaev} and subsequently studied by several authors \cite{Guinea, Fisher&Zwerger, Schoen,Froehlich&Zegarlinski}. 

\subsection{Single electron box}

The so-called single electron box, which consists of a low-capacitance metallic island connected to an outside lead by a tunnel junction, is described by the action of a dissipative quantum rotor. Due to the large charging energy of the island, the presence of excess charges influences single electron tunneling and such a device thus exhibits Coulomb blockade phenomena.   

\begin{figure}[t]
\centering
\includegraphics [angle=0, width=0.4\textwidth] {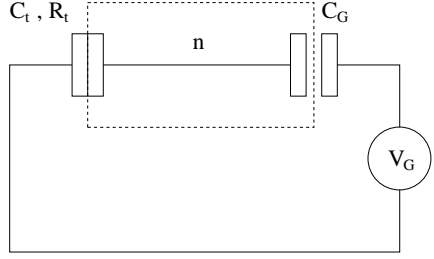}
\caption{Circuit diagram of the single electron box. The box with excess charge $n$ is indicated by the dashed rectangle. It is connected to a voltage source through a capacitor $C_G$ and a tunnel junction with resistance $R_t$ and capacitance $C_t$.}.
\label{box}
\end{figure}

A circuit diagram of a single electron box is shown in Fig.~\ref{box}. The box with excess charge $n$ is controlled by an external voltage source $V_G$, to which it is connected through a capacitor $C_G$ and a tunnel junction with resistance $R_t$ and capacitance $C_t$. The bare charging energy
$E_C=\frac{e^2}{2(C_t+C_G)}$ sets the energy scale. An applied gate voltage $V_G$ induces a continuous polarization charge $n_G=C_GV_G/e$. 

The effective action of the single electron box with gate charge zero can be derived from a microscopic theory, as detailed in Refs. \citen{diss, Schoen, Hofstetter_diplom, Ben-Jacob, Eckern}: 
\begin{eqnarray}
S[\theta] &=& \frac{1}{4E_C} \int_0^\beta d\tau \dot\theta(\tau)^2 - \frac{1}{2\pi e^2 R_t}\int_0^\beta d\tau d\tau' \frac{(\pi/\beta)^2\cos(\theta(\tau')-\theta(\tau))}{(\sin((\pi/\beta)(\tau-\tau')))^2}. 
\label{action_ebox_intro}
\end{eqnarray}
The compact angular variable $\theta$  is conjugate to the number of excess charges on the island and
the partition function 
can be written as the sum over all paths with winding number $\omega=\pm n$, $n=0,1,2\ldots$.

At low temperatures,  $k_BT \ll E_C$, and high tunneling resistance, $R_t \gg R_K=h/e^2\approx 25.8 k\Omega$, the number of excess charges on the island is a staircase function with unit jumps at $n_G=1/2$ (mod 1).    
Thermal fluctuations will round off the corners of the step structure. But even at zero temperature, electron tunneling processes can smear out the staircase, which for large tunneling approaches a straight line. These quantum fluctuations renormalize the ground state energy and lead to an effective charging energy $E^*_C$, which
can be computed from the expectation value of the winding number $\omega$ squared \cite{Hofstetter},
\begin{equation}
\frac{E_C^*}{E_C}=\frac{2\pi^2}{\beta E_C}\langle \omega^2\rangle = \frac{2\pi^2}{\beta E_C}\frac{1}{Z}\int_0^{2\pi} d\theta_0 \sum_{n=0}^\infty n^2\int_{\theta_0}^{\theta_0\pm 2\pi n}\mathcal{D}\theta e^{-S[\theta]}.
\label{winding}
\end{equation}
In chapter 4 we will address the behavior of $E^*_C/E_C$ in the limit of large tunneling conductance. 
\section{Algorithms}

In this main section we present algorithms used for the simulation of the dissipation coupled quantum systems discussed in the previous chapter. A brief introduction to the Monte Carlo method and the concept of importance sampling will be followed by a section on cluster updates for spin systems and the efficient treatment of long-range interactions. Cluster algorithms identify clusters of spins which can be updated simultaneously in a rejection-free move. Near phase transitions, where domains of aligned spins occur on all length scales,  they dramatically reduce autocorrelation times (the number of updates needed to produce an uncorrelated configuration) compared to local update schemes. 

\subsection{Importance sampling and the Metropolis Algorithm}

 We would like to compute the expectation value of some observable $O$,
\begin{equation}
\langle O \rangle = \sum_\Omega P(x)O(x),
\label{expectation}
\end{equation} 
where $P$ denotes a probability distribution defined on the configuration space $\Omega$. Since it is usually not possible to sum over all configurations, the strategy is to sample random configurations $x_1,\ldots, x_N$, according to their importance $P(x_i)$ and approximate Eq.~(\ref{expectation}) by
\begin{equation}
\overline O \approx \frac{1}{N}\sum_{i=1}^{N} O(x_i).
\label{approx_expectation}
\end{equation} 
For statistically independent $O(x_i)$ and large enough $N$ it follows from the central limit theorem that $\overline O$ follows a Gaussian distribution centered at $\langle O \rangle$ and standard deviation
\begin{equation}
\sigma \equiv \Delta \overline O = \sqrt{\frac{\text{Var(O)}}{N}}\approx \sqrt{\frac{\overline{O^2}-{\overline O}^2}{N-1}}.
\label{delta_O}
\end{equation}
As a consequence, the accuracy of a Monte Carlo simulation scales as $1/\sqrt{N}$ with computation time $N$.

The challenge is to generate a chain of configurations $x_1, \dots, x_N$ such that the $x_i$ are sampled according to their weight $P(x_i)$. For this purpose we introduce a Markov process which defines a probability distribution $P_{t+1}$ at time (Monte Carlo step) $t+1$ from the distribution $P_t$ at time $t$ by means of a transition matrix $T$,
\begin{equation}
P_{t+1}(y) = \sum_x T_{y|x}P_t(x).
\label{markov}
\end{equation} 
The elements $T_{y|x}$ of $T$ correspond to the transition probabilities from state $x$ to state $y$ and therefore
\begin{equation}
 T_{y|x}\ge0, \hspace{5mm}\sum_y T_{y|x}=1.
\label{positivity}
\end{equation}
As a consequence $P_{t+1}$ is a probability distribution if $P_t$ is one. 
An additional requirement for $T$ is ergodicity: it must be possible to reach every configuration $y$ from any configuration $x$ in a finite number of Markov steps. This assures that in the limit of infinitely many steps, the whole configuration space is sampled. Finally, $T$ must be such that the probability distribution $P$ is \textit{stationary},
\begin{equation}
P(y) = \sum_x T_{y|x}P(x).
\label{stationarity}
\end{equation}  
This condition is satisfied, if 
\begin{equation}
T_{y|x}P(x)=T_{x|y}P(y),
\label{detailed_balance}
\end{equation}
as is obvious by summing over $x$. The relation (\ref{detailed_balance}) is referred to as \textit{detailed balance}. 

Given the above, any probability distribution $P_0(x)=\delta_{x, x_i}$ will converge to $P(x)$ as $t\rightarrow\infty$ \cite{kemeny_snell}. Hence, an infinitely long Markov chain starting with $x_i$ and defined by the transition matrix $T$ generates configurations according to their weight $P(x)$.   

We would now like to explicitly construct the transition matrix $T$. To this end we decompose the transition probability to move from a configuration $x$ to a configuration $y$, $T_{y|x}$, into an `a priori' probability $p^\text{prop}_{y|x}$ to propose the move from $x$ to $y$ and  $p^\text{acc}_{y|x}$ to accept the move. Assuming that $p^\text{prop}_{x|x}=0$ and $p^\text{prop}_{y|x}$ satisfies conditions (\ref{positivity}), the transition matrix defined as
\begin{equation}
T_{y|x} = \left\{
\begin{array}{ll}
p^\text{prop}_{y|x}p^\text{acc}_{y|x} & y\ne x\\
1-\sum_{y\ne x} p^\text{prop}_{y|x}p^\text{acc}_{y|x} & y=x
\end{array}
\right.
\end{equation}
satisfies Eq.~(\ref{positivity}) as well. The detailed balance condition (\ref{detailed_balance}) now reads
\begin{equation}
p^\text{prop}_{y|x}p^\text{acc}_{y|x} P(x) = p^\text{prop}_{x|y}p^\text{acc}_{x|y} P(y), 
\end{equation}
and can be satisfied in several ways. The Metropolis algorithm \cite{Metropolis} is based on the choice
\begin{equation}
p^\text{acc}_{y|x} = \min\left(1, \frac{p^\text{prop}_{x|y} P(y)}{p^\text{prop}_{y|x} P(x)}\right). 
\label{metropolis}
\end{equation}

\subsection{Cluster algorithms for classical spins}

The Monte Carlo simulation of spin systems near criticality requires the use of efficient algorithms, which we will first discuss for the Ising model with nearest neighbor interactions ($K>0$, $\langle i, j\rangle$ denotes a pair of nearest neighbor sites)
\begin{equation}
S = -K\sum_{\langle i, j\rangle}\sigma_i \sigma_j.
\label{ising}
\end{equation}
Since the correlation length diverges in the vicinity of a phase transition,
large domains of aligned spins appear. Thus, local update schemes become very inefficient in changing the overall configuration, which leads to large autocorrelation times. In order to overcome this problem, 
which means to induce transitions between probable configurations which differ on large scales,
it is necessary to flip entire
clusters of spins
in each Monte Carlo
update. The Swendsen-Wang algorithm \cite{Swendsen&Wang} chooses these clusters in such
a way, that detailed balance is automatically satisfied (no cluster move is rejected). 

\begin{figure}[t]
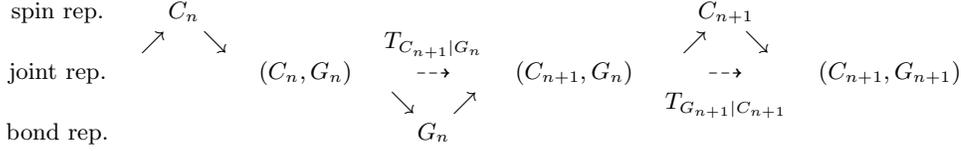

\centering
\begin{tabular}{ccccccccc}
spin rep. 	& $C_n$ &			&  &  &$C_{n+1}$		\\
				& $\nearrow\hspace{5mm}\searrow$  & &$T_{C_{n+1}| G_n}$ & &$\nearrow\hspace{5mm}\searrow$ &  \\
joint rep. & & $(C_n,G_n)$ & $\dashrightarrow$ & $(C_{n+1}, G_n)$ & $\dashrightarrow$ & $(C_{n+1}, G_{n+1})$\\
				& &  & $\searrow\hspace{5mm}\nearrow$ &  & $T_{G_{n+1}| C_{n+1}}$ \\
bond rep. & & & $G_n$ &&&\\
\end{tabular}
\caption{Illustration of the Markov process in the Swendsen-Wang cluster algorithm, which switches back and forth between spin configurations ($C_n$) and bond configurations ($G_n$). The transition probabilities $T_{C_{n+1}| G_n}$ and $T_{G_{n+1}| C_{n+1}}$ can be found from the joint representation in terms of spin and bond variables.}
\label{markov_illustration}
\end{figure}

The Swendsen-Wang algorithm is an example of a dual Monte Carlo algorithm \cite{Kandel&Domany}, which switches back and forth between two configuration spaces, as illustrated in Fig.~\ref{markov_illustration}. One representation is in terms of Ising \textit{spin variables} $\sigma_i\in\{-1, 1\}$, the other one is the random cluster representation of Fortuin and Kasteleyn \cite{Fortuin&Kasteleyn}, which is expressed in terms of \textit{bond variables} $b_{ij}\in \{0,1\}$. The transition probabilities between configurations in both spaces follow directly from the joint Edwards-Sokal representation in terms of spin and bond variables \cite{Edwards&Sokal, Evertz_high_temperature}. 

The partition function of the Ising model (\ref{ising}) in these three representations reads
\begin{itemize}
\item spin representation
\begin{equation}
Z = \sum_{\{\sigma\}}e^{K\sum_{\langle i, j \rangle}\sigma_i \sigma_j} =  \sum_{\{\sigma\}} \prod_{\langle i, j \rangle} e^{K\sigma_i\sigma_j}, 
\label{partition_spin}
\end{equation}
\item joint representation
\begin{equation}
Z = \sum_{\{b\}}\sum_{\{\sigma\}}\prod_{\langle i, j \rangle}e^K\Big[\delta_{b_{ij},0}e^{-2K}+\delta_{b_{ij},1}\delta_{\sigma_i, \sigma_j}(1-e^{-2K})\Big],
\label{partition_joint}
\end{equation}
\item bond representation
\begin{equation}
Z = e^{-K N_b^\text{tot}} \sum_{\{b\}}(e^{2K}-1)^{n_b} 2^{n_c},
\label{partition_cluster}
\end{equation}
where $N_b^\text{tot}$ denotes the total number of bonds, $n_b$ the number of bond variables with value $b_{ij}=1$ and $n_c$ the number of clusters.
\end{itemize}
The joint representation in Eq.~(\ref{partition_joint}) is obtained from Eq.~(\ref{partition_spin}) by
introducing bond variables $b_{ij}\in\{0,1\}$ using the identity
\begin{equation}
e^{-K}e^{K\sigma_i\sigma_j} = \sum_{b_{ij}}\Big[\delta_{b_{ij},0}e^{-2K}+\delta_{b_{ij},1}\delta_{\sigma_i, \sigma_j}(1-e^{-2K})\Big].
\label{identity}
\end{equation}
Summing over the spin variables in Eq.~(\ref{partition_joint}) then yields the Fortuin-Kasteleyn random cluster representation, Eq.~(\ref{partition_cluster}). 

The Swendsen-Wang algorithm follows directly from the joint representation~\cite{Edwards&Sokal}. Given a spin configuration $\{\sigma_i\}$, the conditional probability for freezing a bond ($b_{ij}=1$) is found from Eq.~(\ref{partition_joint}) to be
\begin{eqnarray}
T_{b_{ij} = 1|\sigma_i=\sigma_j} &=& \frac{1-e^{-2K}}{e^{-2K}+ (1-e^{-2K})} = 1-e^{-2K},\\
T_{b_{ij} = 1|\sigma_i\ne\sigma_j} &=& 0.
\end{eqnarray}
If the bond configuration $\{b_{ij}\}$ is given, we find
\begin{eqnarray}
T_{\sigma_i=\sigma_j | b_{ij} = 1} &=& 1,\\
T_{\sigma_i=\sigma_j | b_{ij} = 0} &=& 1/2,
\end{eqnarray}
which means that all spins of a given cluster point in the same direction, whereas the spins on different clusters are uncorrelated after the update.

In the more general case, where spins interact over arbitrary distances with different strengths, the probability of a bond between the sites $i$ and $j$ becomes
\begin{equation}
P_\text{bond}(i,j) = \max(0,1-\exp[-\Delta S(i,j)]),
\label{p_bond}
\end{equation}
where 
$\Delta S(i,j) = S_\text{updated}(i,j)-S_\text{original}(i,j)$
denotes the cost in action 
of flipping a spin associated with the pair ($i,j$).

\begin{figure}[htbp]
\begin{minipage}[b]{0.43\linewidth}
\centering\epsfig{figure=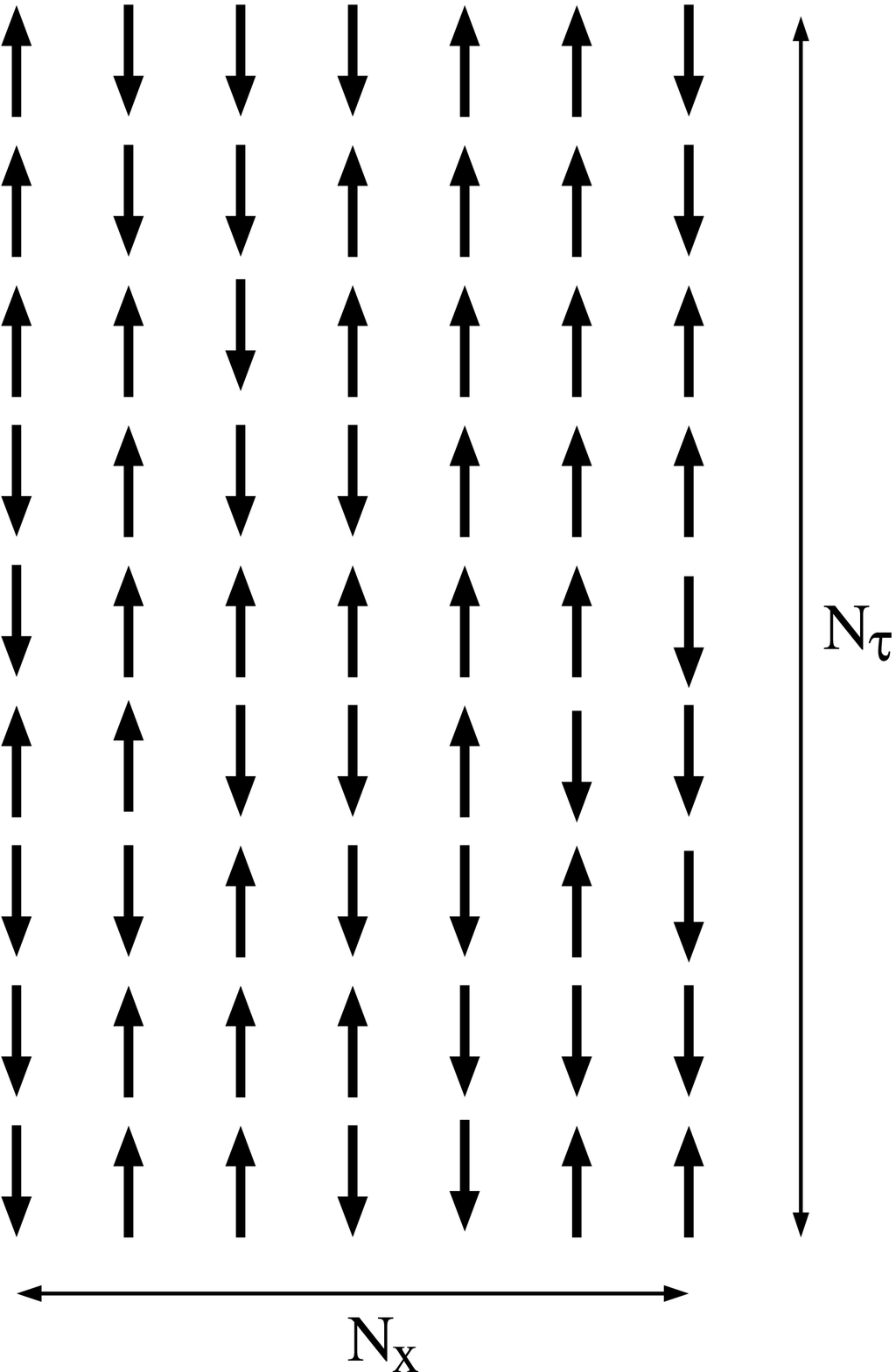,angle=0,width=\linewidth}
\end{minipage} \hfill
\begin{minipage}[b]{0.43\linewidth}
\centering\epsfig{figure=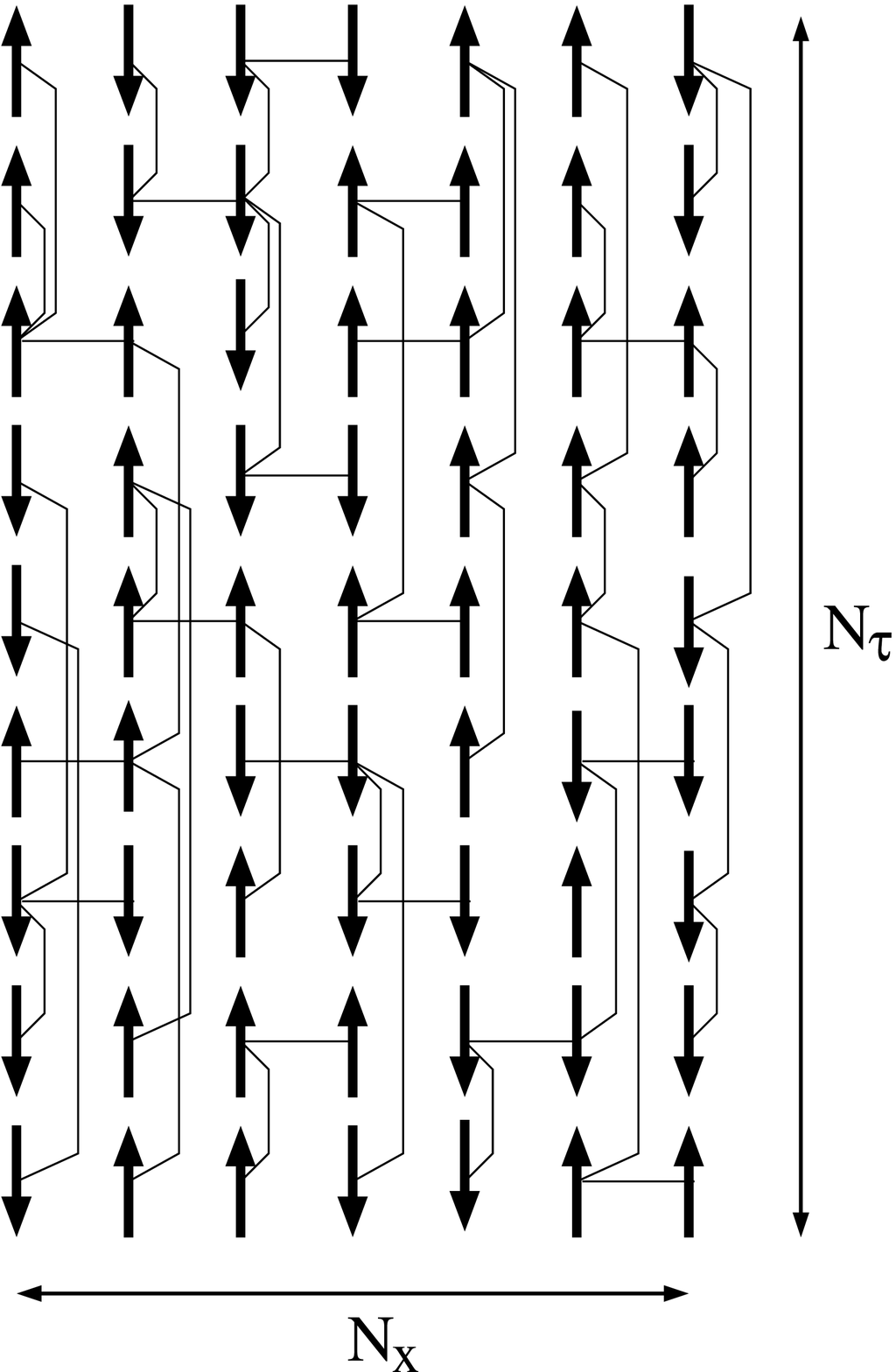,angle=0,width=\linewidth}
\end{minipage}\\ \vspace{1mm}\\
\begin{minipage}[b]{0.43\linewidth}
\centering\epsfig{figure=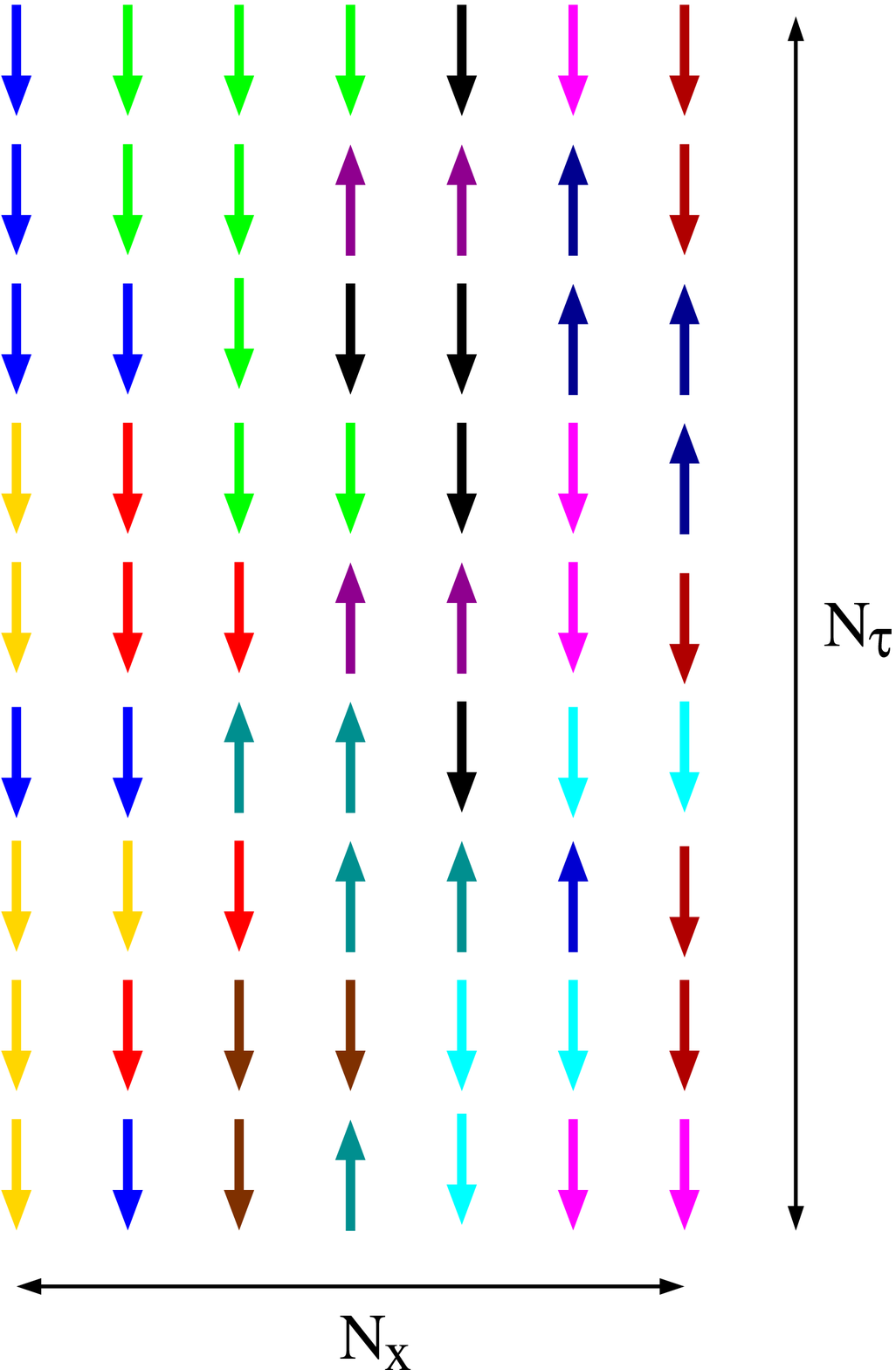,angle=0,width=\linewidth}
\end{minipage} \hfill
\begin{minipage}[b]{0.43\linewidth}
\centering\epsfig{figure=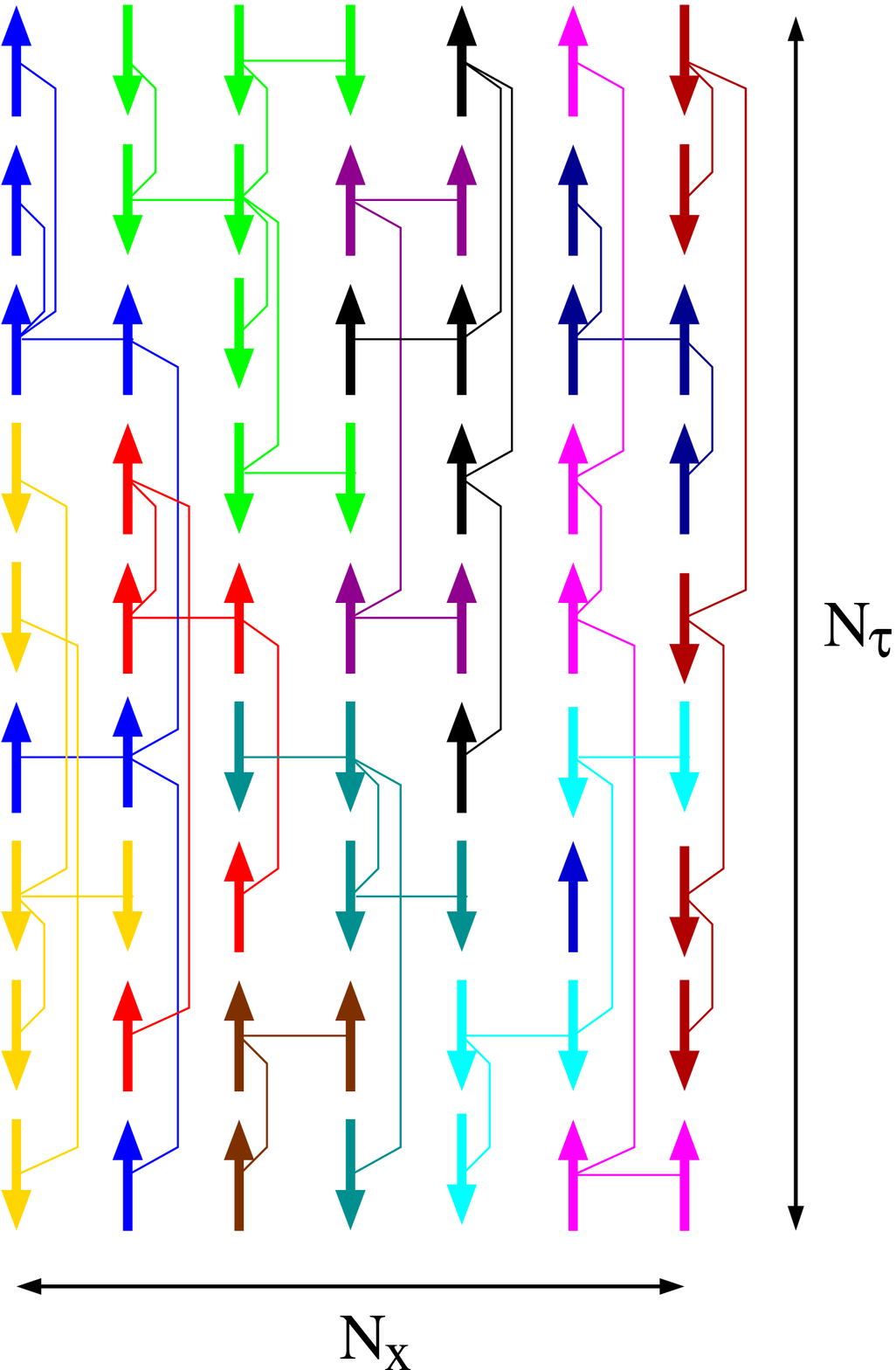,angle=0,width=\linewidth}
\end{minipage}
\caption{Illustration of the Swendsen-Wang cluster algorithm. The figures show clockwise from the top left: (i) original spin configuration, (ii) insertion of bonds,
(iii) clusters of connected spins, (iv) new spin configuration obtained by flipping the spins of each cluster with probability $\frac{1}{2}$.}
\label{cluster}
\end{figure}

The Swendsen-Wang algorithm proceeds in the following three steps, which for the case of the dissipative quantum Ising chain (\ref{spin_lattice}) are illustrated in Fig.~\ref{cluster}
\begin{enumerate}
\item Bonds are inserted between the sites of interacting spins with probability (\ref{p_bond}).
\item Clusters of connected spins are identified.
\item The spins of each cluster are flipped with probability $\frac{1}{2}$. 
\end {enumerate}

Wolff \cite{Wolff} extended the ideas of Swendsen and Wang to $O(n)$ spin models. First of all he noted that it is more efficient to build a single cluster from a randomly chosen site and update all the spins in this cluster. This has the advantage that most updates take place in the largest regions of aligned spins. By choosing a random direction $\vec e$ on the $n$-sphere and projecting the spins onto this direction, one ends up with an Ising like system. Updating a spin then means flipping this projected component, or in other words mirroring the spin on the plane perpendicular to $\vec e$
\begin{equation}
\vec\sigma \rightarrow \vec\sigma - 2 (\vec\sigma\cdot \vec e) \vec e.
\label{4.19}
\end{equation}
The probability (\ref{p_bond}) for inserting a bond between two spins at sites $i$ and $j$ remains valid. A Wolff cluster update therefore proceeds in the following three steps
\begin{enumerate}
\item Choose a random site and a direction $\vec e$.
\item Find all spins connected to this site using the bond probability (\ref{p_bond}), where $\Delta S(i,j) = S(\sigma_i - 2 (\vec\sigma_i\cdot \vec e) \vec e, \vec \sigma_j) - S(\vec \sigma_i, \vec \sigma_j)$
\item Mirror all the spins in the cluster on the plane perpendicular to $\vec e$.
\end{enumerate}

\subsection{Efficient treatment of long-range interactions}

In the presence of long-range forces, every site interacts with every other site and an algorithm which
iterates over all bonds would be of order $O(N^2)$, where $N$ is
the number sites. This is prohibitively slow and thus we present here two more sophisticated
algorithms, proposed by Luijten and Bl\"ote \cite{Luijten&Bloete}, which are of order $O(N\log N)$ and $O(N)$ respectively. These ideas are essential for the simulation of dissipative quantum systems at low temperature, since we have seen in chapter 2 that dissipation introduces long-range interactions in the imaginary time direction.

We consider first a chain of classical Ising spins with long-range interactions,
\begin{equation}
S = -\sum_{i,j}g(i-j)\sigma_i\sigma_j.
\label{ising_long}
\end{equation}
The kernel $g$ of interest in our applications is of the form ($j\ne0$)
\begin{equation}
g(j)=\alpha\frac{(\frac{\pi}{N})^2}{(\sin(\frac{\pi}{N}j))^2},
\label{kernel_general}
\end{equation}
but this is of no importance for the following discussion.
From Eq.~(\ref{p_bond}) it follows that the probability
of a long-range bond between two parallel spins at sites 0 and $j$ is
\begin{equation}
P_{\text{bond}}(0;j) = 1 - e^{-2g(j)}.\label{5.1}
\end{equation}
Hence, the probability that no bond is formed between spin 0 and
spins $j+1,\ldots,n-1$ (all assumed parallel) becomes
\begin{equation}
P_{\text{no bond}}(0;j+1,\ldots,n-1) =
\prod_{i=j+1}^{n-1}e^{-2g(i)}
=\exp\Bigg[-2\sum_{i=j+1}^{n-1}g(i)\Bigg].\label{5.3}
\end{equation}
If we define an array (``lookup-table") $A$ of length $N$ with
elements $A[0]=1$, 
and $A[n] =\prod_{i=1}^{n}e^{-2g(i)}$ for $n=1,\ldots,N-1$,
then Eq.~(\ref{5.3}) can be written as
\begin{equation}
P_{\text{no bond}}(0;j+1,\ldots,n-1)=\frac{A[n-1]}{A[j]}.\label{5.5}
\end{equation}
The spins connected to site 0 are then calculated as follows: a
random number $r_1$ is chosen in the interval $[0,1)$ and the
array $A$ searched for the first index, say $n_1$, such that
$A[n_1]\le r_1$. For the calculation of the next bond, the values
of $A$ must be divided by $A[n_1]$, or equivalently, the random
number $r_2$ multiplied by $A[n_1]$. The site number $n_2$ of the
second connected spin is the first index, such that $A[n_2]\le
A[n_1]r_2$. This process is continued until
$A[n_{k-1}]r_k < A[N-1]$. 

The outlined procedure assumes that all spins are aligned. Before actually inserting a proposed bond, we have to check whether the two spins are parallel.  If the bisection method is used to search for the lowest index
such that $A[n]\le r$, the calculation time is of order $O(\log N)$ per site. An update of the whole lattice with lookup-table is therefore of
order $O(N\log N)$.

The search of a lookup-table could even be avoided, if it is possible to evaluate Eq.~(\ref{5.3}) analytically, as is the case for the kernel (\ref{kernel_general}). To this end we approximate the sum in the exponent by an integral
\begin{equation}
\sum_{i=j+1}^{n-1}K_i \approx
\alpha\int_{j+\frac{1}{2}}^{n-\frac{1}{2}} d\tau\frac{(\frac{\pi}{N})^2}{(\sin(\frac{\pi}{N}\tau))^2}
=-\alpha\frac{\pi}{N}\left[\cot\Big(\frac{\pi}{N}\Big(n-\frac{1}{2}\Big)\Big)-\cot\Big(\frac{\pi}{N}\Big(j+\frac{1}{2}\Big)\Big)\right].\label{5.6}
\end{equation}
Given the value of $j$ (index of the previous bond), the value of
$n$ (index of the next bond) is calculated from a uniformly
distributed random number $r\in [0,1)$ by inverting
\begin{equation}
\exp\Bigg[-2\sum_{i=j+1}^{n-1}K_i\Bigg] = 1-r,\label{5.7}
\end{equation}
where it is understood that $n$ will be truncated to the next
smaller integer (because the sum is only up to $n-1$). Replacing the sum in Eq.~(\ref{5.7}) by Eq.~(\ref{5.6}) we get
\begin{equation}
n =
\text{int}\left[\frac{1}{2}+\frac{N}{\pi}\text{atan2}\left(1,\frac{1}{\tan(\frac{\pi}{N}(j+\frac{1}{2}))}+\frac{N}{\pi}\frac{\ln
(1-r) }{2\alpha}\right)\right].\label{5.9}
\end{equation}
The function atan2 satisfies $\tan(\text{atan2}(x,y))=x/y$. It is defined on the interval $[-\pi,\pi]$ and the
signs of the two arguments are used to determine the quadrant. 

If the analytical formula (\ref{5.9}) is used to calculate bonds,
the computation time per site is of order $O(1)$. In fact, from Eqs.~(\ref{5.3}) and
(\ref{5.6}) it follows with $j=0$ and $n=N$,
\begin{equation}
P_{\text{no bond}} = \text{exp}\left\{2\alpha\frac{\pi}{N}\left[\cot\Big(\pi-\frac{\pi}{2N}\Big)-\cot\Big(\frac{\pi}{2N}\Big)\right]\right\}\approx e^{-2\alpha},\label{5.10}
\end{equation}
where in the last step we assumed $N$ large. Hence, irrespective of the value of $N$, the probability for the algorithm to terminate at a given step is at least $e^{-2\alpha}$ and thus the expected number of bonds is finite.

Since short-range interactions are not well approximated by
Eq.~(\ref{5.6}), non-universal quantities, such as critical couplings, will have different values in this analytical
implementation. However, universal quantities, such as critical
exponents, will be calculated correctly, as they only depend on
the asymptotic behavior of the interaction, which is accurately
reproduced by Eq.~(\ref{5.6}).

Although the algorithm with analytical determination of bonded sites scales as $O(N)$ and the algorithm based on a lookup table only as $O(N\log N)$, it turns out that the latter is faster by a factor of 2 for accessible system sizes. 

In the case of $O(n)$-spins, where the bond strength depends on the spin components projected on some random direction $\vec{e}$,
\begin{equation}
\sigma^{\text{proj}} = \vec \sigma\cdot\vec e,
\label{proj}
\end{equation}
one proposes bonds as explained for the Ising case. 
To account for the fact that the interaction is weaker than assumed in the look-up table, a random number $r\in[0,1)$ is chosen and
a proposed bond between sites $i$ and $j$
is only inserted if
\begin{equation}
r(1-e^{-2g(i-j)})<1-e^{-2g(i-j)\sigma_i^{\text{proj}}\sigma_j^{\text{proj}}}.
\label{xy_bond}
\end{equation}
In other words, bonds between $O(n)$-spins are inserted with probability $P_\text{bond}^{O(n)}(i,j)$ by proposing bonds with the higher probability $P_\text{bond}^\text{ising}(i,j)$ and accepting them with probability $P_\text{bond}^{O(n)}(i,j)/P_\text{bond}^\text{ising}(i,j)$.

\subsection{Cluster algorithm for resistively shunted Josephson junctions}

In section 2.3 we derived the effective action for a resistively shunted Josephson junction, which in terms of the dimensionless parameter $\alpha=R_Q/R_s$ reads
\begin{equation}
S[\phi] = \int_0^\beta d\tau \left[\frac{1}{16E_C}\Big(\frac{d\phi}{d\tau}\Big)^2-E_J\cos(\phi)\right]
+\frac{\alpha}{8\pi^2}\int_0^\beta \int_0^\beta d\tau d\tau' \frac{(\pi/\beta)^2(\phi(\tau)-\phi(\tau'))^2}{\sin((\pi/\beta)(\tau-\tau'))^2}.
\label{junction_action_method}
\end{equation}  
Because of the Josephson coupling energy and the non-compact nature of the phase variable in this action,
one cannot directly employ the cluster algorithms available for spin systems, which were discussed in the previous section. In order to simulate resistively shunted junctions efficiently, we developed a new algorithm \cite{junction}, consisting of two kinds of updates: (i) local updates in Fourier space compatible with the Gaussian terms in Eq.~(\ref{junction_action_method}) and (ii) rejection-free cluster updates. The first type of moves assures ergodicity of the algorithm and the second type produces global cluster updates compatible with the energetic constraints from the Josephson potential.

\subsubsection{Local updates in Fourier space}

For the Monte Carlo simulation, we discretize imaginary time into $N$ (assumed odd) time steps.
The action (\ref{junction_action_method}) can then be expressed in the simple form 
\begin{equation}
S[\phi] = \sum_{k=0}^{N-1} a_k|\tilde\phi_k|^2-E_J\Delta\tau\sum_{j=0}^{N-1}\cos(\phi_j),
\label{eq3}
\end{equation}  
where $\tilde\phi_k=\sum_{j=0}^{N-1} e^{i\frac{2\pi}{N}jk}\phi_j$ denotes the Fourier transform of $\phi$. The positive coefficients $a_k$ are defined as $a_k= \frac{2}{N}(\tilde g_0-\tilde g_k$), with $\tilde g_k$ the Fourier transform of the kernel ($j\ne 0$)
\begin{equation}
g(j) = \frac{1}{32E_C\Delta\tau}(\delta_{j,1}+\delta_{j,N-1})+\frac{\alpha}{8\pi^2}\frac{(\pi/N)^2}{\sin((\pi/N)j)^2}.
\label{eq4}
\end{equation}
Since $\tilde \phi_k^* = \tilde \phi_{N-k}$, only $\{\tilde\phi_k | k=0,\ldots, (N-1)/2\}$ need to be considered. 

In a local update of the frequency components $\tilde\phi_k$ and $\tilde \phi_{N-k}$, we choose a new value according to the probability distribution of the Gaussian term in Eq.~(\ref{eq3}), 
\begin{equation}
p(\phi_k) \sim e^{-2a_k|\tilde \phi_k| ^2},
\label{eq5}
\end{equation}
by fixing the phase at a random value in the interval $[0, 2\pi]$ and $|\tilde \phi_k|^2$ using exponentially distributed random numbers with mean $1/(2a_k)$. This move is accepted with probability
\begin{equation}
p(\phi_{\text{old}}\rightarrow \phi_{\text{new}}) =
\min\left(1, e^{-\{S_J[\phi_{\text{new}}]-S_J[\phi_{\text{old}}]\}}\right),
\label{eq6}
\end{equation}
where $S_J[\phi]=-E_J\Delta\tau\sum_{j=0}^{N-1}\cos(\phi_{j})$ and $\phi_\text{old}$, $\phi_\text{new}$ denote the backward Fourier transform of the old and new $k$-space configuration, respectively. Such local updates can be performed in a time $O(N)$ and have recently been used in the simulation of 2D Josephson junction arrays \cite{Capriotti}.

\begin{figure}[hp]
\hfill
\begin{minipage}[b]{\linewidth}
\centering
\includegraphics [angle=0, width=0.8\textwidth]{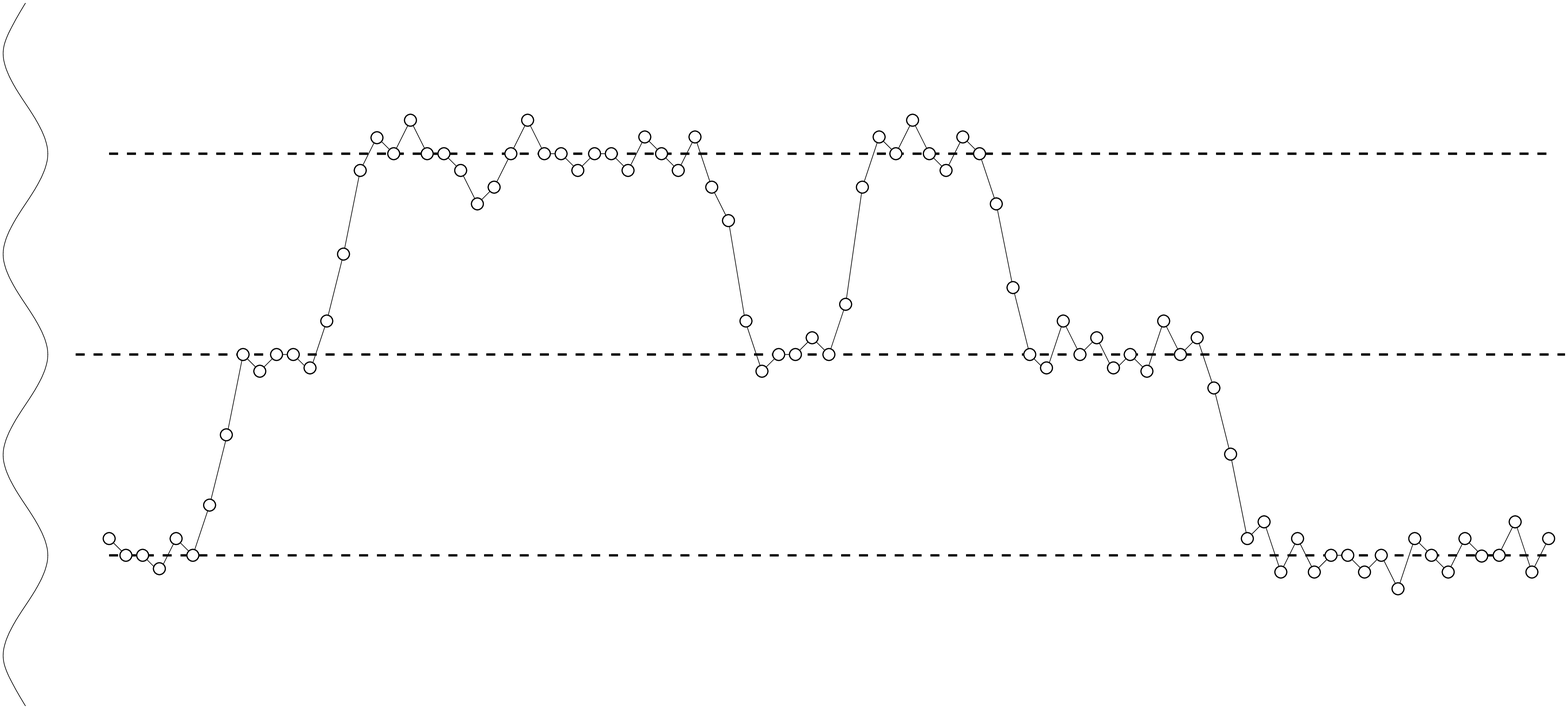}
\end{minipage}
\vspace{1mm}\\
\mbox{}\hfill
\begin{minipage}[b]{\linewidth}
\centering
\includegraphics [angle=0, width=0.8\textwidth]{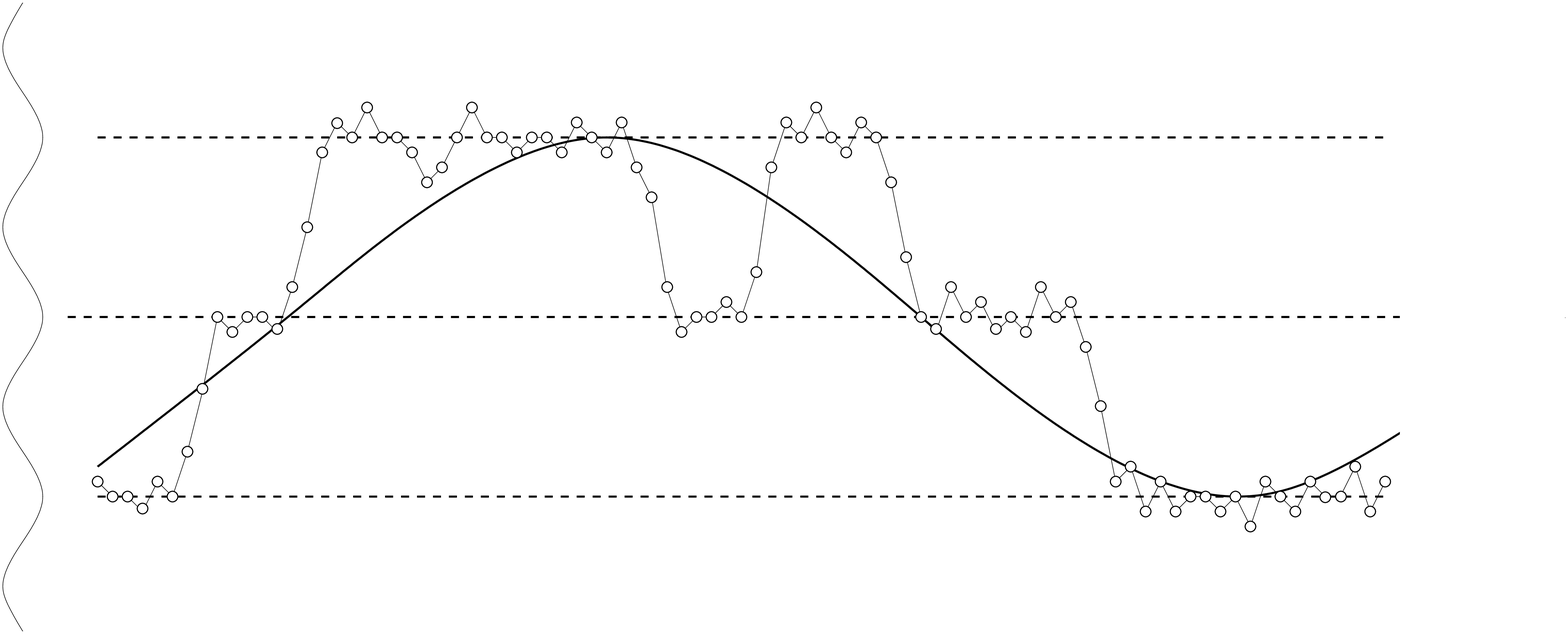}
\end{minipage}
\vspace{1mm}\\
\mbox{}\hfill
\begin{minipage}[b]{\linewidth}
\centering
\includegraphics [angle=0, width=0.8\textwidth]{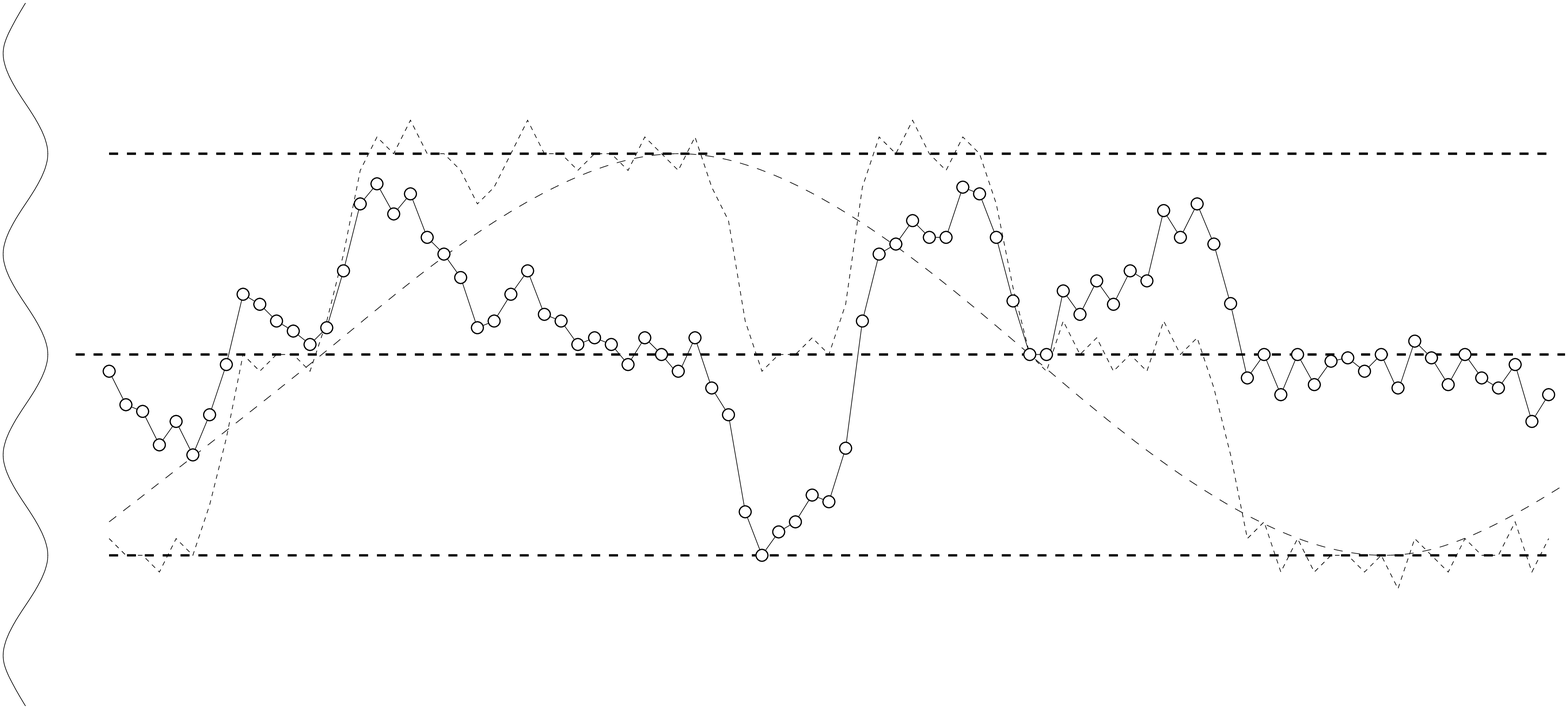}
\end{minipage}
\hfill \mbox{}
\caption{Illustration of local updates in Fourier space. Phase configurations $\{ \phi_j | j=0,\ldots,N-1\}$ are plotted as small circles and the value of the Josephson potential is sketched on the left. Dashed lines indicate the positions where the potential is minimal. 
The figures show from top to bottom: (i) Original phase configuration. The phase variables remain most of the time near a minimum of the cosine-potential, which results in a step-like structure. (ii) Randomly picked Fourier component, for which a new amplitude will be chosen according to Eq.~(\ref{eq5}). (iii) New phase configuration obtained if the new amplitude is close to zero. Since many of the phase variables end up in the region of high potential energy, this move will be rejected with high probability.}
\label{junction_local}
\end{figure}

\subsubsection{Cluster updates}

\begin{figure}[hp]
\hfill
\begin{minipage}[b]{\linewidth}
\centering
\includegraphics [angle=0, width=0.8\textwidth]{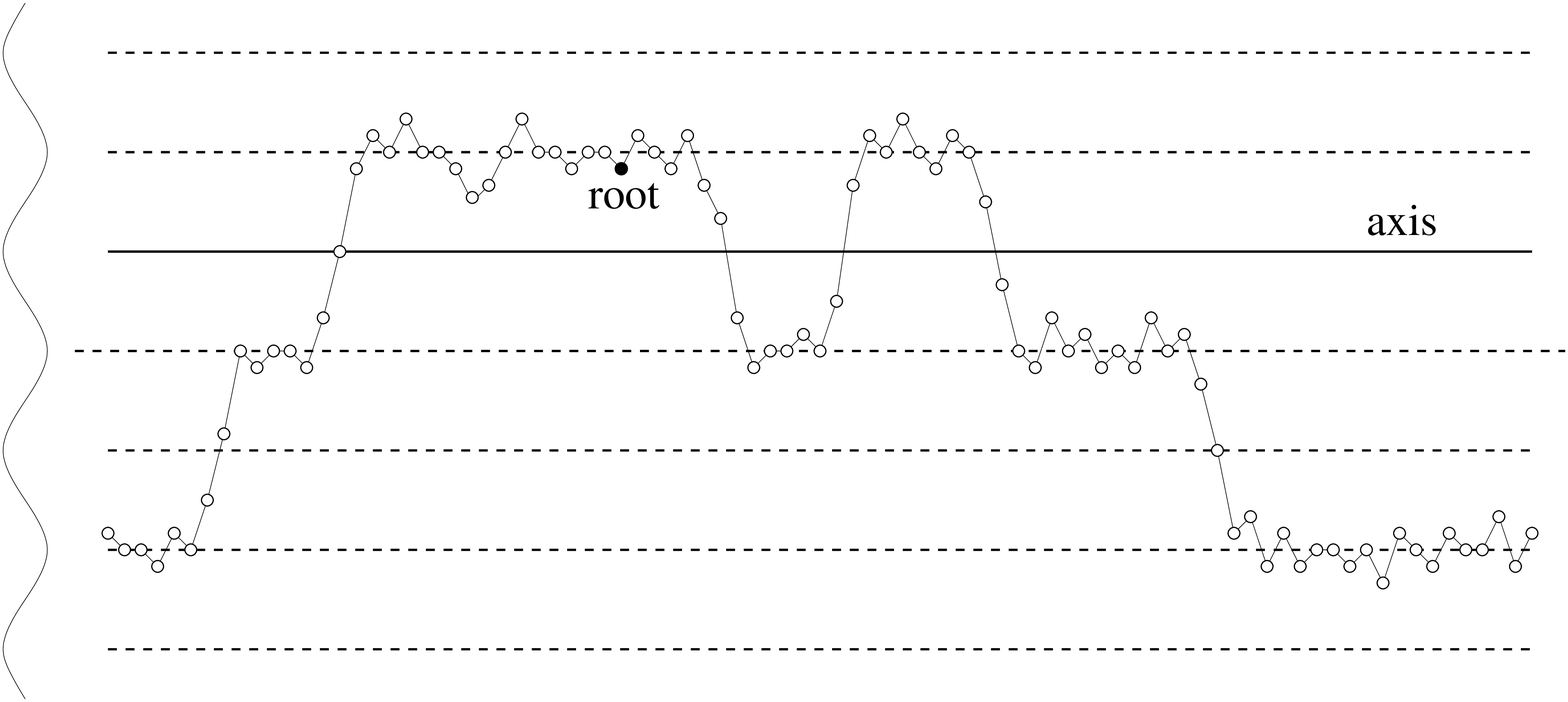}
\end{minipage}
\vspace{1mm}\\
\mbox{}\hfill
\begin{minipage}[b]{\linewidth}
\centering
\includegraphics [angle=0, width=0.8\textwidth]{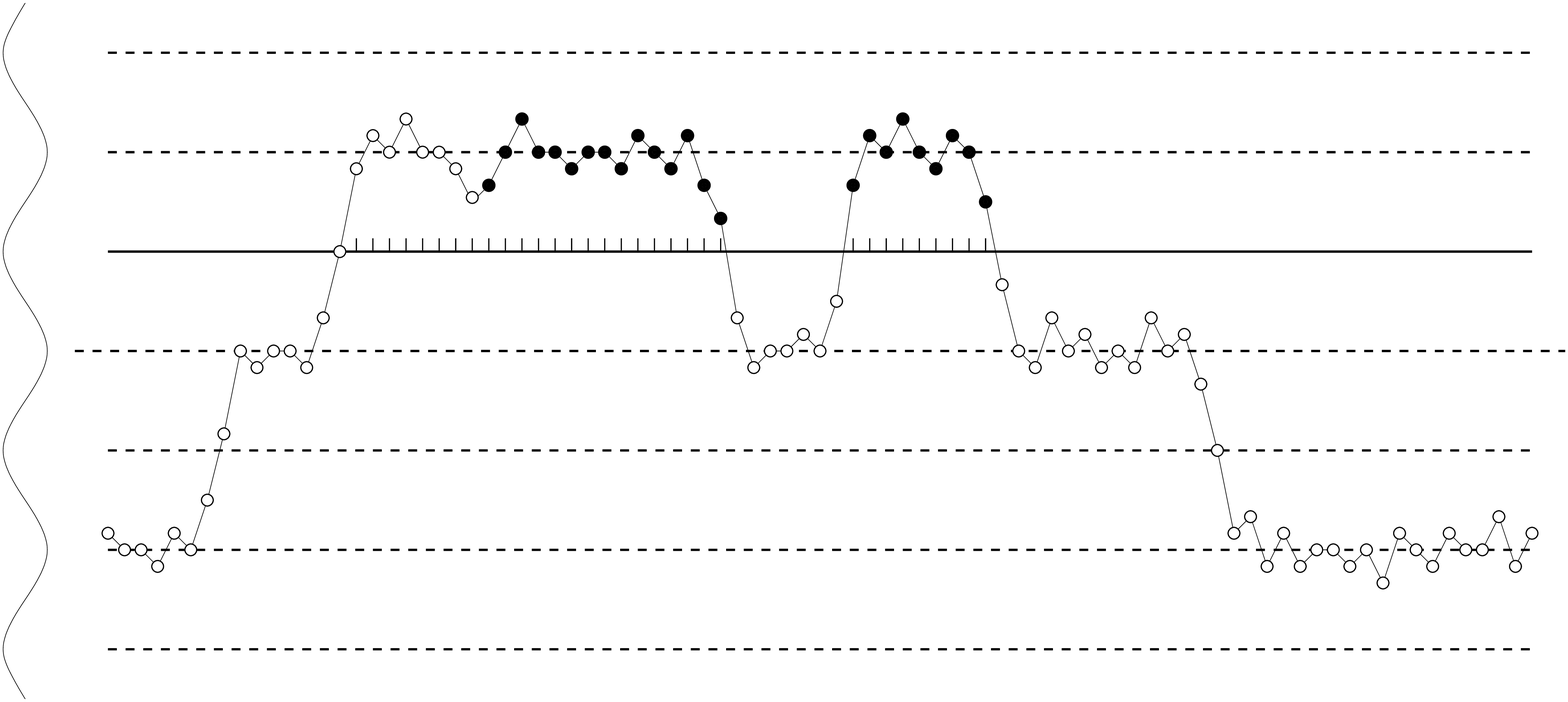}
\end{minipage}
\vspace{1mm}\\
\mbox{}\hfill
\begin{minipage}[b]{\linewidth}
\centering
\includegraphics [angle=0, width=0.8\textwidth]{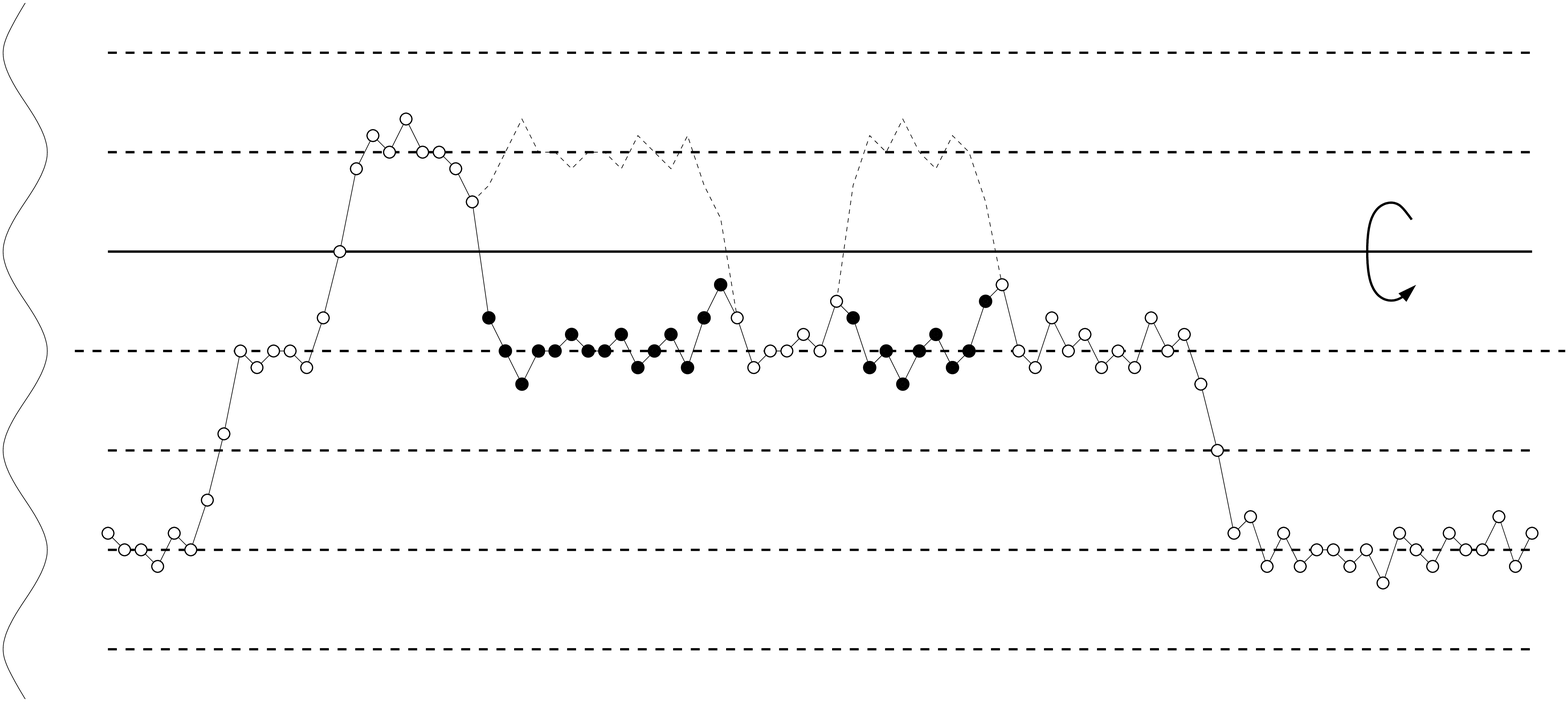}
\end{minipage}
\hfill \mbox{}
\caption{Illustration of the cluster algorithm for non-compact phase variables. The figures show from top to bottom: (i) Original phase configuration. Possible axis positions are indicated with dashed lines, located at the maxima and minima of the cosine potential. The randomly chosen axis and root site of the cluster are marked with the black solid line and black dot respectively. (ii) Cluster of connected sites. The sites which could potentially connect to the root site are indicated with tic marks. (iii) New phase configuration obtained by flipping the cluster around the axis.}
\label{junction_cluster}
\end{figure}

For reasonably large values of $E_J$, local $k$-space updates which introduce phase changes on the order of $2\pi$ will be strongly suppressed, because their sinusoidal shape does not resemble an optimal phase slip path (see illustration in Fig. \ref{junction_local}). Algorithms based on local updates alone will therefore be ineffective near the phase transition, where phase slips start to proliferate. A typical path will stay most of the time near one or the other of the minima in the cosine-potential, as shown in Fig. \ref{junction_cluster}. 
A simple idea for a global update compatible with this overall structure would be step-updates which shift the phases $\phi_j$ by $\pm 2\pi$ in some random interval $[j_\text{min}, j_\text{max}]$. It would, however, be better to let the algorithm choose itself the phase variables $\phi_j$, which can be shifted to a different potential minimum.

The observation which leads to such a cluster algorithm is that a shift by multiples of $\pm 2\pi$ is not the only operation which leaves $S_J$ invariant. The same is true for reflections on $\{n\pi\}_{n\in \mathrm{Z}}$, that is, the positions of the maxima and minima of the cosine potential. 
We exploit the latter symmetry to design a rejection-free cluster update consisting of the following four steps, which are illustrated in Fig.~\ref{junction_cluster}:

\begin{enumerate}
\item An axis $\phi=n^\text{axis}\pi$ with integral $n^\text{axis}$ in the range $[-n_\text{max}, n_\text{max}]$ is randomly chosen (the significance of $n_\text{max}$ is discussed below) and a random site $j_\text{root}$ is picked as the root site of the cluster. We introduce relative coordinates
\begin{equation}
\phi^\text{axis}_j = \phi_j - n^\text{axis}\pi,
\label{eq8}
\end{equation}
which are updated in a cluster move as $\phi_j^\text{axis}\rightarrow -\phi_j^\text{axis}$, in complete analogy to the projected spin components (\ref{proj}) in the Wolff algorithm. Such updates do not change the value of $S_J$. 
\item The cluster of sites connected to $j_\text{root}$ is constructed in a way analogous to the case of $O(n)$-spins which was explained in section 3.2.
Two sites at positions $i$ and $j$ are connected with probability
\begin{equation}
p(i,j) = \max\Big(0, 1-e^{-\{S[\phi^\text{axis}_i,-\phi^\text{axis}_j]-S[\phi^\text{axis}_i,\phi^\text{axis}_j]\}}\Big),
\label{eq9}
\end{equation}
where
\begin{equation}
S[\phi^\text{axis}_i,-\phi^\text{axis}_j]-S[\phi^\text{axis}_i,\phi^\text{axis}_j] = 8g(i-j)\phi^\text{axis}_i \phi^\text{axis}_j.
\label{10}
\end{equation}
This expression is (up to a factor of two) identical to the spin case, because the quadratic contributions cancel.

\item The phases of the sites $j$ belonging to the cluster are updated according to 
\begin{equation}
\phi^\text{axis}\rightarrow -\phi^\text{axis}.
\label{update_phi}
\end{equation}

\item If necessary, the whole configuration is shifted by multiples of $\pm 2\pi$ in such a way, that the mean value $\bar\phi=\frac{1}{N}\sum_{j=0}^{N-1}\phi_j$ is closest to the potential minimum corresponding to $n^\text{axis}=0$. The last step is important to assure detailed balance with a finite $n_\text{max}$. The parameter $n_\text{max}$ must be large enough that the re-centered configuration is contained in the interval $[-n_\text{max}\pi, n_\text{max}\pi]$. This value can be determined during the thermalization of the system.
\end {enumerate}

There are alternatives to fixing the interval of symmetry points and re-centering the new configurations, which obviously satisfy detailed balance. One such possibility would be to choose the axis among the $n_\text{max}$ symmetry points above or below $\phi_{j_\text{root}}$. Another idea, proposed in Ref.~\citen{Evertz}, is to choose the axis closest to the phase variable of a randomly chosen site. This has the advantage that the axis cuts through the phase configuration with high probability, which means that few freezing clusters (cluster which span the entire system) in the delocalized phase are produced.  

\begin{figure}[t]
\centering
\includegraphics [angle=-90, width= 0.7\textwidth] {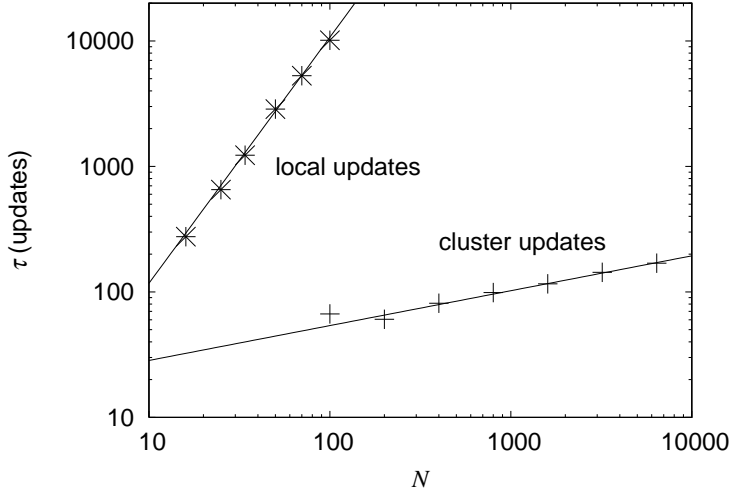}
\caption{Integrated autocorrelation times $\tau$ for $\langle (\phi-\bar\phi)^2\rangle$ versus system size $N$. The data were obtained at the critical point $\alpha=1$ using $E_J/E_C=1$.}
\label{auto}
\end{figure}

Using the ideas of Luijten and Bl\"ote \cite{Luijten&Bloete} discussed in section 3.2.3, the cluster move can be performed in a time $O(N\log N)$ despite long-range interactions, which allows us to compute precise data for large systems (up to $N\approx 10^4$ if $\Delta\tau E_C = 0.25$). Fig.~\ref{auto} plots the integrated autocorrelation time $\tau$ for $\langle (\phi-\bar\phi)^2 \rangle$ as a function of system size $N$. Even though the CPU time for a cluster update is $O(N\log N)$ as compared to $O(N)$ for a local update, the gain in sampling efficiency is considerable.

\subsection{Winding number sampling}

In section 2.4 it was shown that the effective charging energy of the single electron box can be computed from the winding number distribution of the phase configurations. If we discretize imaginary time into $N$ time steps  $\Delta\tau=\beta/N$, the action (\ref{action_ebox_intro}) becomes
\begin{equation}
S[\phi]=- \frac{1}{2E_C\Delta\tau}\sum_{j=1}^{N}\cos(\phi_{j+1}-\phi_j)-\alpha\sum_{j<j'}\frac{(\frac{\pi}{N})^2 \cos(\phi_j-\phi_{j'})}{\sin^2(\frac{\pi}{N}(j-j'))},
\label{discreteaction}
\end{equation} 
where we have introduced the dimensionless tunneling conductance $\alpha = \frac{1}{2\pi^2}\frac{R_K}{R_t}$, with $R_K=\frac{h}{e^2}$ ($\alpha = \frac{1}{\pi e^2 R_t}$ for $\hbar=1$).
Periodic boundary conditions $\phi_{N+1}=\phi_1$ are employed. Since only $2\pi$-periodic functions appear in Eq.~(\ref{discreteaction}), $\phi_j \in [0,2\pi)$ can be interpreted as the angle which defines the orientation of an XY-spin. Comparing Eqs.~(\ref{spin_lattice}) and (\ref{discreteaction}) we see that the action of the single electron box is equivalent to that of a single dissipative quantum rotor.  

In Fourier space, Eq.~(\ref{discreteaction}) becomes local,
\begin{equation}
S[\phi]=\frac{1}{N}\sum_{k=0}^{N-1}g_k|\psi_k|^2,
\label{fourier_action}
\end{equation} 
where $\psi_k=\sum_{j=0}^{N-1}e^{i\frac{2\pi}{N}jk}e^{i\phi_j}$ denotes the Fourier transform of $e^{i\phi}$ and $g_k$ the Fourier transform of the kernel ($j\ne 0$)
\begin{equation}
g(j) = -\frac{1}{4E_C\Delta\tau}(\delta_{j,1}+\delta_{j,N-1})-\frac{\alpha}{2}\frac{(\pi/N)^2}{\sin((\pi/N)j)^2}.
\label{kernel}
\end{equation}
Using fast Fourier transformation it is therefore possible to compute the action of a configuration in a time $O(N\log N)$ despite the long-range interactions. This is important for the transition matrix Monte Carlo algorithm presented below. 

\subsubsection{Path integral Monte Carlo}

The expectation value $\langle \omega^2 \rangle$, and hence the effective charging energy (\ref{winding}), can be computed using path-integral Monte Carlo with cluster updates, as explained in section 3.2. This approach works well for small and intermediate values of the tunneling conductance $\alpha$. Configurations with winding number $\omega=\pm n$ are generated according to their weight $w_n$ in the partition sum,
\begin{equation}
w_n=\frac{\int_0^{2\pi}d\phi_0\int_{\phi_0}^{\phi_0\pm2\pi n}\mathcal{D}\phi e^{-S[\phi]}}{\int_0^{2\pi}d\phi_0\sum_{n=0}^{\infty}\int_{\phi_0}^{\phi_0\pm2\pi n}\mathcal{D}\phi e^{-S[\phi]}}.
\label{weight}
\end{equation}
However, for large $\alpha$, the effective charging energy -- and therefore also the fraction of paths with winding number different from zero -- decreases as $\exp(-\pi^2\alpha)$ according to theoretical predictions. A huge number of configurations with winding number zero will be generated for each configuration with non-zero winding number. Due to this rapidly deteriorating efficiency, the cluster algorithm can only be used in the range of tunneling conductances for which $E_C^*\gtrsim 10^{-8} E_C$.

\subsubsection{Transition matrix Monte Carlo}

In order to compute the effective charging energy at even larger values of the tunneling conductance, we developed a new Monte Carlo approach \cite{ebox}, which attempts to insert or remove phase slips (kinks) and in doing so measures the relative weights of the different winding number sectors. As only paths of winding number 0 and 1 contribute in the limit of large $\alpha$, we restrict the discussion to the calculation of the relative weight of these two winding number sectors. However, the algorithm can also be used to study additional winding number sectors.

\begin{figure}[t]
\centering
\includegraphics [angle=0, width= 0.8\textwidth] {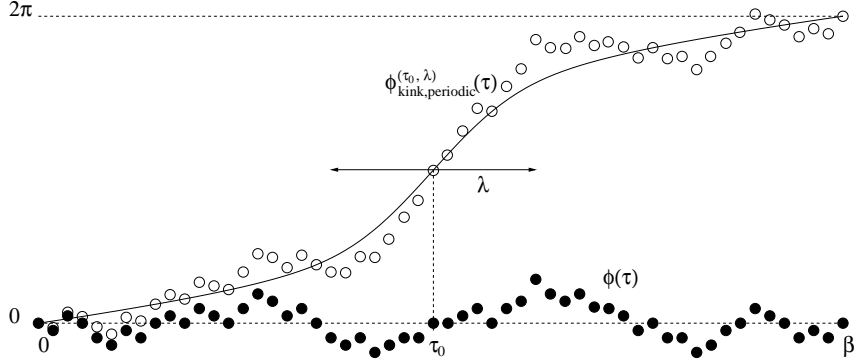}
\caption{Illustration of a kink (phase-slip) insertion. The filled dots show the original phase configuration $\{ \phi_n=\phi(\tau_n) | n=1,\ldots,N\}$ with winding number $\omega=0$, and the line a phase-slip path $\phi_\text{kink, periodic}^{(\tau_0,\lambda)}$. Both the position $\tau_0$ and the width $\lambda$ of this phase slip path are randomly chosen. The open dots show the new configuration with winding number $\omega=1$, which would be obtained by adding $\phi_\text{kink, periodic}^{(\tau_0,\lambda)}$ to $\phi(\tau)$.
}
\label{kink_insertion}
\end{figure} 

The effective charging energy (\ref{winding}) can be expressed for large $\alpha$ as
\begin{equation}
\frac{E_C^*}{E_C} = \frac{2\pi^2}{\beta E_C}\frac{\sum_{n=0}^{\infty} n^2 w_n}{\sum_{n=0}^\infty w_n}\approx\frac{2\pi^2}{\beta E_C}\frac{w_1}{w_0}.
\label{large_winding}
\end{equation}
The weights $w_0$ and $w_1$ are proportional to the time, which a Metropolis-Monte Carlo simulation would spend in winding sectors 0 and 1, respectively. A ``flat histogram", which means equal occupation probabilities for both winding sectors, could be obtained by adding an additional ``inverse-weight" factor to the usual Boltzmann term. 
The acceptance probability for a proposed kink update from a configuration with winding number $n$ and action $S$ to winding number $n'$ and action $S'$ then becomes
\begin{equation}
P_{(n,S)\rightarrow(n',S')}=\min\left(1,\frac{w_n}{w_{n'}}e^{-(S'-S)}\right),
\label{probability_metro}
\end{equation}
whereas a cluster update from $n$ to $n'$ (which takes the change in action into account) would be accepted with probability
\begin{equation}
P_{n\rightarrow n'}=\min\left(1,\frac{w_n}{w_{n'}}\right).
\label{probability_sw}
\end{equation}
The ``flat histogram" (detailed balance) condition for the a priori unknown relative occupation probability $w_1/w_0$ can then be expressed for $n, n' \in\{0,1\}$ as
\begin{equation}
\left\langle P_{0\rightarrow 1}(w_1/w_0)\right\rangle = \left\langle P_{1\rightarrow 0}(w_1/w_0)\right\rangle,
\label{flat_histogram}
\end{equation}
where the averages are taken over a random sequence of kink- and cluster updates.

Our strategy is to determine the transition probabilities $\left\langle P_{0\rightarrow 1}\right\rangle$ and $\left\langle P_{1\rightarrow 0}\right\rangle$ for different values of $w_1/w_0$ in order to find the solution of Eq.~(\ref{flat_histogram}), rather than trying to adjust the relative occupation of the winding sectors by a Wang-Landau type iterative procedure \cite{Wang&Landau}. This is an approach in the spirit of the transition Matrix Monte Carlo method \cite{Wang&Swendsen}.

Each winding number sector is treated separately and the cluster updates serve essentially to randomize the configurations within that sector, although their contribution to the probabilities $\langle P_{i\rightarrow j}\rangle$ becomes important for smaller $\alpha$. In a kink update, illustrated in Fig.~\ref{kink_insertion}, we insert a phase slip  $\phi_{\text{kink}}^{(\tau_0, \lambda)} = \pm 2\arctan\left((\tau-\tau_0)/\lambda\right)$ of random width $\lambda$ at some random position $\tau_0$, or rather 
\begin{equation}
\phi_{\text{kink, periodic}}^{(\tau_0, \lambda)} = \pm \pi\frac{\arctan((\tau-\tau_0)/\lambda)}{\arctan(\beta/2\lambda)},
\label{kink_periodic}
\end{equation}
periodically continued outside the interval $-\frac{\beta}{2}\le \tau-\tau_0\le\frac{\beta}{2}$, which is a slightly modified version compatible with the finite size of the system. These are stationary paths of the long-range part of the action (\ref{action_ebox_intro}) in the limit $\beta\rightarrow\infty$.
\begin{figure}[t]
\centering
\includegraphics [angle=-90, width= 0.7\textwidth] {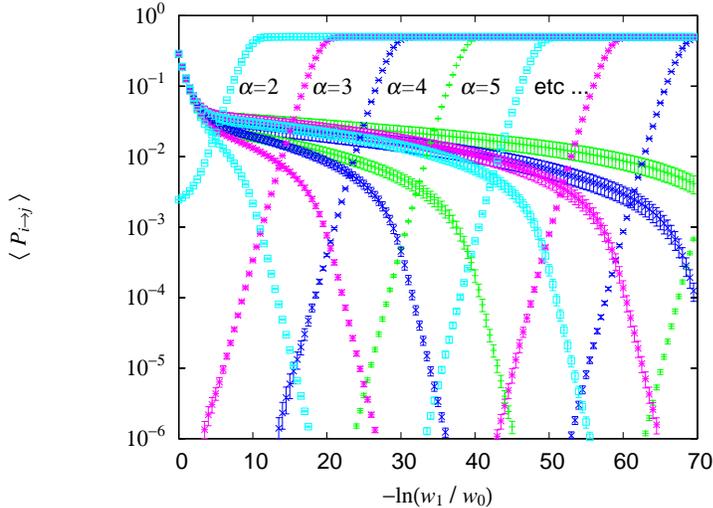}
\caption{Transition probabilities $\langle P_{0\rightarrow 1}\rangle$ (with positive slope) and $\langle P_{1\rightarrow 0}\rangle$ (with negative slope) plotted as a function of $-\ln(w_1/w_0)$. The intersection points of these curves determines $w_1/w_0$. From left to right, the data correspond to $\alpha=2$, 3, $\ldots$, 9.}
\label{intersect}
\end{figure} 

Inserting $\phi_{\text{kink}}^{(\tau_0, \lambda)}$ changes the original configuration $\phi$ with winding number $n$ to $\phi'$ with winding number $n\pm 1$. The corresponding Boltzmann weights $\exp(S[\phi]-S[\phi'])$ are computed using Eq.~(\ref{fourier_action}) and stored during the simulation. Updates which produce a configuration with winding number $|n|>1$ as well as cluster updates which leave the winding number unchanged get the weight 0. A cluster update which connects to the other winding number sector gets the weight 1.  From these data, one can calculate the average transition probabilities $\langle P(i\rightarrow j)\rangle$ as a function of $w_1/w_0$ using Eqs.~(\ref{probability_metro}) and (\ref{probability_sw}). Finally, the relative weight of winding sector~1 is determined by solving Eq.~(\ref{flat_histogram}).

To illustrate this procedure, Fig.~\ref{intersect} shows the probabilities $\langle P(0\rightarrow 1)\rangle$ (with positive slope) and $\langle P(1\rightarrow 0)\rangle$ (with negative slope) as a function of $-\ln(w_1/w_0)$ for several values of $\alpha$. The intersection points of these curves determine $w_1/w_0$. The effective charging energies obtained by this new method are consistent with the results from cluster Monte Carlo simulations for the values of $\alpha$ which can be treated by the latter method. The new approach, however, allows us to simulate the system at much higher tunneling conductance. It increases the range of accessible effective charging energies by 30 orders of magnitude \cite{ebox}.

\section{Applications}

This section summarizes the main results of our efforts to simulate environmentally coupled quantum systems, such as dissipative spin chains or resistively shunted Josephson junctions.

In Ref.~\citen{ising}, we studied the phase transition in a chain of Ising spins, using the cluster algorithm detailed in sections 3.2 and 3.3.  Since the two spin states could represent for example the two flux states in a SQUID, such a chain can be thought of as a model of a qubit-register. 
Our careful investigation showed, that no locally critical phase exists even for large dissipation strength, and that the second order phase transition from the disordered to the ordered state is controlled by a single fixed point. In particular, the dynamical critical exponent takes the value $z\approx 2$. This finding, as well as the values of the critical exponents $\eta$ and $\nu$, were in remarkably good agreement with the predictions from a dissipative $\phi^4$-field theory \cite{Pankov, smt}.    

A similar agreement with analytical results was found for the chain of dissipative XY-spins. This chain was studied in Ref.~\citen{smt} as a model for a nano-wire formed out of superconducting grains. Besides the bulk phase transition, we studied the properties of finite, open chains. If normal leads are attached, the (infinite) wire becomes insulating, while superconducting leads turn the whole device superconducting. For mixed leads (normal and superconducting), the wire is metallic and its zero-bias conductance \textit{universal}, 
in the sense that it does not matter how strongly the superconducting lead fixes the phase of the neighboring grain.

In Ref.~\citen{ebox}, we have tested the new algorithm described in section 3.5.2 and found that it allows to calculate the effective charging energy $E_C^*$ up to large values of the tunneling conductance $\alpha$. The leading exponential suppression, $E_C^*/E_C\sim \exp(-\pi^2\alpha)$, can be traced over more then 34 orders of magnitude -- compared to 2 (8) orders of magnitude for path-integral Monte Carlo simulations with local (cluster) updates. In the zero-temperature regime, we measured a pre-exponential factor $\sim\alpha^5$, which is not consistent with any of the numerous theoretical predictions \cite{Hofstetter, Wang_Grabert, Panyukov_Zaikin, Lukyanov, Koenig}. 

With the powerful cluster algorithm described in section 3.4 and Ref.~\citen{junction}, we were able to verify that the superconductor-to-metal transition in a single resistively shunted Josephson junction occurs for $R_s=R_Q$, independent of the Josephson coupling $E_J$. Furthermore, the temperature dependence of the resistance in the $T=0$ superconducting phase is proportional to $T^{2(R_Q/R_s-1)}$, as predicted in Ref.~\citen{Korshunov1}. Remarkably, on the phase transition line, we found continuously varying correlation exponents. The exponent $\eta(q)$, defined through $\langle \exp(iq(\phi_\tau-\phi_0))\rangle \sim \tau^{-2\eta(q)}$ (where $q$ is some non-integral real number), turned out to decrease exponentially with increasing  Josephson energy. Hence, it appears that the superconductor-to-metal transition is in fact a line of fixed points with continuously varying exponents. 

In the two-junction model discussed in Ref.~\citen{Refael1}, such a line of fixed points was predicted to control the superconductor-to-metal transition. Our numerical calculations in Ref.~\citen{twojunction}, which utilize adapted versions of the cluster moves discussed in section 3.4, showed, that the critical properties in this interacting system are, within error-bars, the same as in the single junction and a mean field theory (which assumes non-interacting junctions) was even capable of accurately predicting the phase boundary for intermediate Josephson coupling. These examples illustrate how the ability to simulate single junctions and extended systems with high precision has led to new insights and will be important in future investigations.

\subsection*{Acknowledgments}
The research reported here was supported by the Swiss National Science Foundation, the Kavli Institute for Theoretical Physics and the Aspen Center for Physics. MT thanks the Yukawa International Seminar 2004 for the hospitality in Kyoto.

%

\end{document}